\documentclass[aps,prd,showpacs,floatfix,twocolumn,10pt]{revtex4-1}
\usepackage[utf8]{inputenc}
\usepackage[T1]{fontenc}
\usepackage{graphicx}
\usepackage{amsmath}
\usepackage{amsfonts}
\usepackage{braket}

\RequirePackage[colorlinks=true, urlcolor=blue, anchorcolor=blue, citecolor=blue, filecolor=blue, linkcolor=blue, menucolor=blue, linktocpage=true, pdfproducer=medialab, pdfa=true]{hyperref}

\usepackage{bm}
\newcommand{\hypgf}{{}_2{F}_1}
\newcommand{\lp}{\ell_\mathrm{P}}
\newcommand{\FI}{\mathcal{F}}

\usepackage{txfonts}

\let\Im\relax

\DeclareMathOperator{\Im}{Im}
\DeclareMathOperator{\Tr}{Tr}
\DeclareMathOperator{\var}{Var}
\begin{document}
\title{Probing deformed quantum commutators}
\author{Matteo A. C. Rossi}
\email{matteo.rossi@unimi.it}
\homepage{http://users.unimi.it/aqm}
\affiliation{Quantum Technology Lab, 
Dipartimento di Fisica, Universit\`a degli Studi di Milano, 
20133 Milano, Italy}
\author{Tommaso Giani}
\email{tommaso.giani@studenti.unimi.it}
\affiliation{Dipartimento di Fisica, Universit\`a degli Studi di Milano, 
20133 Milano, Italy}
\author{Matteo G. A. Paris}
\email{matteo.paris@fisica.unimi.it}
\homepage{http://users.unimi.it/aqm}
\affiliation{Quantum Technology Lab, Dipartimento di Fisica, 
Universit\`a degli Studi di Milano, 20133 Milano, Italy}
\affiliation{INFN, Sezione di Milano, I-20133 Milano, Italy}
\date{\today}
\begin{abstract}
Several quantum gravity theories predict a minimal length at the order of magnitude of the Planck length, under 
which the concepts of space and time lose their physical meaning. In quantum mechanics, the insurgence of 
such minimal length can be described by introducing a modified position-momentum commutator, which in turn 
yields a generalized uncertainty principle (GUP), where the uncertainty on position measurements has a lower bound. 
The value of the minimal length is not predicted by theories and must be estimated experimentally. In this paper, 
we address the quantum bound to estimability of the minimal uncertainty length by performing measurements on 
a harmonic oscillator, which is analytically solvable in the deformed algebra induced by the deformed commutation 
relations. 
\end{abstract}

\maketitle
\section{Introduction}
The existence of a minimal length is a general feature of many quantum gravity theories (see \cite{Garay1995,Hossenfelder2013} and references therein).
According to these theories, the Planck length
\begin{equation}
	\lp=\sqrt{\hbar G/c^3} \simeq 1.6 \cdot 10^{-35}\,m	
\end{equation}
sets an order of magnitude under which the concepts of space and time lose their physical meaning. 
In turn, this corresponds to the existence of a minimal uncertainty in a position measurements, which sets 
a limit to the localizability of an object.
\par
The uncertainty principle derived from the standard commutation relations between position and momentum
 does not predict the existence of any inferior bound to the position uncertainty, as the latter may be arbitrary 
 small, provided that momentum uncertainty gets bigger. From this fact derives the idea of modifying the 
 commutation relation between position and momentum, in order to obtain the prediction of a minimal position 
 uncertainty \cite{Maggiore1993,Kempf1994,Kempf1995,Kempf1997,Milburn2006}.
\par
In one dimension, let us consider the minimal deformation 
\begin{equation}\label{eq:modified_commutator}
\left[\bm{x},\bm{p} \right]=i\hbar\left[1+\beta_0 \left(\frac{\lp p}{\hbar}\right)^2\right],
\end{equation}
$\beta_0$ being a positive dimensionless parameter. It is easy to see that the following generalized 
uncertainty principle (GUP) holds
\begin{equation}\label{eq:gup}
\Delta x\Delta p \geq \frac{\hbar}{2}\left[1+\beta_0\left(\frac{\lp\Delta p}{\hbar}\right)^2\right].
\end{equation}
Equation \eqref{eq:gup} does indeed predict an inferior bound to position uncertainty, given by
 $\Delta x_0= \lp\sqrt{\beta_0}$.
\par
The introduction of a deformed commutator as in Eq. \eqref{eq:modified_commutator}, 
modifies the algebra of the Hilbert space and alters the spectral decomposition of the Hamiltonian 
operator of many quantum systems of theoretical and experimental interest. Among them, the 
harmonic oscillator is of paramount theoretical importance and several studies have been focused 
on it, in the context of deformed commutators \cite{Kempf1995,Lewis2011,Chang2002}. The energy 
eigenvalues can be found analytically in an arbitrary number of dimensions and the eigenstates in 
the momentum basis can be obtained \cite{Kempf1995,Chang2002}.

The value of $\beta_0$ in Eqs. \eqref{eq:modified_commutator} and \eqref{eq:gup}, usually assumed 
to be around unit \cite{Das2008}, has to be found experimentally since theoretical predictions are still lacking.
Recently, beside proposed tests with high-energy or neutrino experiments \cite{Harbach2004,Sprenger2011}, an opto-mechanical experimental scheme has been proposed \cite{Pikovski2012}, and an upper 
 bound to the value of $\beta_0$ has been set in \cite{Bawaj2015}, using micro- and nano-mechanical 
 harmonic oscillators. Since $\beta_0$ does not correspond to a proper quantum observable, its value 
 should be inferred through some indirect measurements, which causes an additional error in its estimation. 
 In particular, if this extra uncertainty is too big compared to the value of the parameter, it may be 
 intrinsically inestimable, and no experiment may be able to observe its presence.

The purpose of this work is to analyze the ultimate limits to precision in the estimation of $\beta_0$, 
exploiting tools from local quantum estimation theory (QET) \cite{Helstrom1976,Holevo2001,Braunstein1994,Paris2009}, 
and presenting the results for an harmonic oscillator prepared in various initial states. Estimation theory provides 
a rigorous framework to determine the bound to the precision achievable in an estimation procedure of 
experimental data. This bound, known as the Cramer-Rao inequality \cite{Cramer1946} is connected to the 
Fisher information of the probability distribution. QET is a generalization to quantum systems: the ultimate 
bound to precision is found by optimizing the Fisher information over all the quantum measurements that can 
be made on a system. By providing the tools to find the optimal measurement and state preparation, QET allows to go beyond standard classical limits in precision and has been successfully applied to a wide range of metrological problems \cite{Giovannetti2004,Giovannetti2011}, in particular in quantum interferometry and quantum optics \cite{Monras2007}, and in experiments with photons \cite{Nagata2007,Berni2015}, trapped ions \cite{Meyer2001,Leibfried2004}.

Remarkably, the study of the modified algebra of the Hilbert space
induced by the deformed commutators has highlighted a shortcoming of
standard QET, that in turn has led us to a critical revision and
generalization of the standard Cram\'er-Rao bounds \cite{Seveso2016},
which we will discuss in the following.  We also notice that deformation
of position commutators also occurs in other models, e.g. due to spin
induced uncertainty \cite{der14}, and the corresponding effects may be
observable at different length scales. 

The paper is structured as follows. In Section \ref{sec:harmonic_oscillator} we report the solution of the 
eigenvalues problem for the harmonic oscillator in the modified algebra, reporting explicit expressions 
for the energy spectrum and for the eigenfunctions. In Section \ref{sec:qet} we review some results of local QET, 
reporting the expression for Fisher information (FI), quantum Fisher information (QFI) and estimability of a parameter. 
In Section \ref{sec:qfi_harmonic_oscillator} we present the main results of our work. We discuss the modifications to 
QET required for this problem, and we show the ultimate bounds on precision in the measure of the parameter, 
calculating also the performance of the momentum operator. Analytical expansions for small values of $\beta_0$ 
are derived for FI and QFI relative to pure states. We also analyze the QFI and FI for mixed states and the thermal
 state. Finally, we analyze the dependence of the results on the mass and frequency of the oscillator, in order to 
 find the best experimental configurations.
Section \ref{sec:conclusions} closes the paper with some concluding remarks.

\section{Harmonic oscillator}
\label{sec:harmonic_oscillator}
In this Section we consider the linear harmonic oscillator in the algebra generated by $\bm{x}$ and $\bm{p}$ obeying the commutation relation
\begin{equation}
[\bm{x},\bm{p}]=i\hbar(1+\beta p^2),
\label{eq:commutator}
\end{equation}
with $\beta=\lp^2/\hbar^2\beta_0$, which has the units of inverse square momentum.

The action of position and momentum as differential operators in the momentum representation is given by
\begin{align}
\bm{p}\,\psi(p) & = p\, \psi(p) \label{eq:momentum_op_def} \\
\bm{x}\,\psi(p) & = i\hbar(1+\beta p^2)\partial_p\psi(p). \label{eq:position_op_def}
\end{align}
For the operators $\bm{x}$ and $\bm{p}$ to be symmetric, and thus represent physical observables, the scalar product of the Hilbert space must be modified:
\begin{align}
\langle\psi|\phi\rangle & =\int_{-\infty}^{+\infty} dp \mu_\beta(p)\psi^*(p)\phi(p)
\label{eq:scalar_product_def} \\
1 & =\int_{-\infty}^{+\infty}{dp}\mu_\beta(p)|p\rangle\langle p|. \label{eq:identity_resolution}
\end{align}
where
\begin{equation}\label{eq:integration_measure}
	\mu_\beta(p) = \frac1{(1+\beta p^2)}.
\end{equation}
The presence of the non-trivial integration measure $\mu_\beta(p)$ has a remarkable impact on the estimatibility of $\beta$, as we will explain in the following Section. 

The Hamiltonian of the harmonic oscillator,
\begin{equation}
\mathcal{H} = \frac{\bm{p}^2}{2 m}+m\omega^2\frac{\bm{x}^2}{2},\label{eq:ho_hamiltonian}
\end{equation}
leads to the following stationary Schr\"{o}dinger equation in the momentum representation: 
\begin{equation}\label{eq:harmonic_oscillator_equation}
\left[-\frac{\hbar^2 k}{2}\left({(1+\beta p^2)\frac{\partial}{\partial p}}\right)^2+\frac{p^2}{2m}\right]\psi(p)= E \psi(p), 
\end{equation}
where $k = m \omega^2$.

The solution of Eq. \eqref{eq:harmonic_oscillator_equation} has been addressed in \cite{Kempf1995} and, i
n a different way, in \cite{Chang2002}. In the former, the
solutions are found, using the general theory of totally
Fuchsian equations, in terms of the hypergeometric function $\hypgf(a,b;c;z)$, while in the latter it is given in terms of the Gegenbauer polynomials $C_n^{(\lambda)}(s)$. 
The solutions of \cite{Kempf1995} and \cite{Chang2002} in the momentum basis are,
respectively,
\begin{widetext}
\begin{align}
\psi_n(p) & = \mathcal{N}_n(1+\beta^2)^{-\frac{1}{2}(n + \lambda)}  \hypgf\left(-n,1-n-2\lambda;1-n-\lambda;\frac{1}{2}\left(1+ip\sqrt{\beta}\right)\right) \label{eq:kempf}  \\
& = \frac{\sqrt[4]{\beta}2^{\lambda-\frac{1}{2}}}{\sqrt{\pi}}\Gamma(\lambda)\sqrt{\frac{n!(\lambda+n)}{\Gamma(n+2\lambda)}} (1+\beta p^2)^{-\frac{\lambda}{2}}  C_n^{(\lambda)}\left(p\sqrt{\frac{\beta}{1+\beta p^2}}\right), \label{eq:chang}
\end{align}
\end{widetext}
where $\lambda=\frac{1}{2}\left\{1+\sqrt{1+4/[(\hbar m\omega)^2\beta^2]}\right\}$
and $\mathcal{N}_n$ is a normalization constant. The relation between
these two solutions involves transformation formulas for the hypergeometric functions. Besides, in \cite{Kempf1995}
the normalization constant $\mathcal{N}_n$ of Eq. \eqref{eq:kempf} is not derived explicitly. The two solutions are compared in Appendix \ref{app:solutions}, where the normalization constant is found to be
\begin{equation}
\mathcal{N}_n=\frac{(-i)^n\sqrt{\pi}\sqrt[4]{\beta}2^{\lambda+n-\frac{1}{2}}}{\sin(\pi\lambda)\Gamma(1-n-\lambda)}\sqrt{\frac{\lambda+n}{n!\Gamma(n+2\lambda)}}.
\end{equation} 

The energy eigenvalues, according to Refs. \cite{Kempf1995,Chang2002,Lewis2011}, are
\begin{equation}
E_n  = \frac{k}{2} \left[\left(n+\frac{1}{2}\right)\left(\Delta x_0^2 + \sqrt{\Delta x_0^4 + 4a^4}\right) + \Delta x_0^2n^2\right],
\label{eq:eigenvalues}
\end{equation}
with $\Delta x_0  = \hbar \sqrt{\beta}$ and $a = \sqrt{\frac{\hbar}{m\omega}}$.

\section{Local Quantum Estimation Theory}
\label{sec:qet}
The parameter $\beta$ introduced in the commutator, Eq. \eqref{eq:commutator}, does not correspond to a proper quantum observable and it cannot be  measured directly.
In order to get information about $\beta$, we have to resort to indirect measurements, inferring its value by the 
measurements of a different observable or a set of observables, that is, we have a parameter estimation problem.

Quantum estimation theory (QET) provides tools to find the optimal measurement according to some given criterion.
In this context we exploit {\itshape local} QET which looks for the quantum measurement that maximizes the so-called {\itshape Fisher information } i.e. minimizing the variance of the estimator at a fixed value of the parameter.
Our aim is to evaluate the ultimate bound on precision, i.e. the smallest value of the parameter that can be discriminated, and to determine the optimal measurement achieving these bounds. 

In the following, we briefly review the main concepts of local QET and set the notation for the rest of the paper. We 
refer the reader to \cite{Paris2009} for a more detailed review of the subject. In the following Section we also discuss 
the generalization of standard QET that is required in the problem at hand, in which the geometry of the Hilbert space 
is affected by the minimal length, i.e. by the parameter to be estimated.

In order to solve an estimation problem we have to find an estimator, i.e. a map from the set of measurements $x_1,x_2,\ldots,x_n$ into the space of parameters $\beta$:
\begin{equation}
	\widehat{\beta}=\widehat{\beta}(x_1,x_2,\ldots,x_n).
\end{equation}

Optimal estimators are those saturating the Cram\'er-Rao inequality \cite{Cramer1946}
\begin{equation}
\var(\beta)\geq\frac{1}{M F(\beta)},
\label{eq:Cram\'er_rao}
\end{equation}
which sets a lower bound on the variance $\var(\beta)=E_{\beta}[(\widehat{\beta}({x})-\beta)^2]$ of any estimator.
$M$ is the number of measurements and $F(\beta)$ is the Fisher information, defined by
\begin{equation}
F(\beta)=\int dx\,P(x|\beta)\left(\partial_\beta \ln\,P(x|\beta)\right)^2,
\label{eq:fi_definition}
\end{equation}
where $P(x|\beta)$ is the probability of obtaining the value $x$ when the parameter has the value $\beta$ and $\partial_\beta$ is a shorthand for $\frac{\partial}{\partial \beta}$.

In quantum mechanics, we consider a {\itshape quantum statistical model} i.e. a family of quantum states 
$\rho_\beta$ defined on a Hilbert space {\itshape H} and labeled by the parameter $\beta$ which in our 
problem is real and positive. 
We want to estimate its value through the measurement of some observable on the state $\rho_\beta$.
A quantum estimator for the parameter $\beta$ is a  pair, 
consisting of a positive-operator valued measurement (POVM) and a classical estimator that accounts for the post-processing of the sampled data. The choice of the quantum measurement is the central problem of QET, since different
choices in general lead to different attainable precisions.

In quantum mechanics the probability of a certain outcome is given by the Born rule $P(x|\beta)=\Tr[{\Pi_x}\rho_\beta]$,
where ${\Pi_x}$, are the elements of the POVM we measure and satisfy $\int dx\,\Pi_x=1$. The FI is then written 
\begin{equation}\label{eq:FI_for_POVM}
  F(\beta)=\int dx\,\frac{[\partial_\beta\Tr(\Pi_x\rho_\beta)]^2}{\Tr(\Pi_x\rho_\beta)}\;.
\end{equation}

Upon defining the symmetric logarithmic derivative (SLD) $L_\beta$ as the self-adjoint operator satisfying the equation
\begin{equation}
\frac{L_\beta\rho_\beta + \rho_\beta L_\beta}{2}=\frac{\partial \rho_\beta}{\partial\beta},
\label{eq:sld}
\end{equation} 
we have that the FI $F(\beta)$ of any POVM is bounded \cite{Braunstein1994} by the so-called {\itshape Quantum Fisher Information} $H(\beta)$:
\begin{equation}
F(\beta)\leq H(\beta) \equiv \Tr[\rho_\beta{L_\beta}^2]=\Tr[\partial_\beta\rho_\beta L_\beta].
\label{eq:qfi_definition}
\end{equation}
The Cram\'er-Rao inequality now takes the form
\begin{equation}
\var(\beta)\geq\frac{1}{M H(\beta)},
\label{eq:QCR}
\end{equation}
which gives the ultimate bound to precision for any unbiased estimator of $\beta$.

Eq. \eqref{eq:sld} is a Lyapunov matrix equation and a general solution exists. An explicit form for the {\itshape Symmetric Logarithmic Derivative} can be given in the basis in which the density operator is diagonal. Upon writing \begin{equation}
\rho_\beta = \sum_n p_n(\beta) \ket{\psi_n(\beta)}\bra{\psi_n(\beta)},	
\end{equation}
where $\{\ket{\psi_n}\}$ is a complete set in the Hilbert space, we have \cite{Paris2009}
\begin{equation}\label{eq:sld_explicit}
L_\beta=2\sum_{n m}\frac{\langle\psi_m|\partial_\beta\rho_\beta|\psi_n\rangle}{p_n+p_m}|\psi_m\rangle\langle\psi_n|,
\end{equation}
where it is understood that the sum is on the indices for which $p_n + p_m \neq 0$.
Form Eq. \eqref{eq:sld_explicit} follows the explicit formula for the QFI
\begin{equation}\label{eq:qfi_explicit}
H(\beta)=2\sum_{n m}\frac{|\langle\psi_m|\partial_\beta\rho_\beta|\psi_n\rangle|^2}{p_n+p_m}.
\end{equation}

The expression of the QFI gets simpler when we consider a family of pure states described by the wave function ${\psi_\beta}$. In standard quantum mechanics it is straightforward to find that th SLD is $L_\beta = 2\partial_\beta \rho_\beta$ by noticing that $\partial_\beta \rho_\beta = \partial_\beta(\rho_\beta^2) = \partial_\beta \rho_\beta \rho_\beta + \rho_\beta \partial_\beta \rho_\beta$, being $\rho_\beta$ a projector onto the pure state \cite{Paris2009}. This yields 
\begin{equation}\label{eq:qfi_pure_states_standard}
	H(\beta) = 4(\braket{\partial_\beta \psi | \partial_\beta \psi} + \braket{\partial_\beta\psi|\psi}^2).
\end{equation}

From a geometrical perspective, the precision in the estimation of the parameter $\beta$ is 
related to the distinguishability of the corresponding state $\rho_\beta$ from its neighbors. If we 
discriminate between the two values $\beta$ and $\beta+d\beta$, with $d\beta$ infinitesimal, the greater the ``distance'' between $\rho_\beta$ and $\rho_{\beta+d\beta}$, the easier our task will be by making a quantum measurement on the system. Among the different definitions of distance that can be made on the manifold of quantum states, the one that turns out to capture the notion of estimation measure is the Bures distance \cite{Bures1969,Uhlmann1976}, defined as
\begin{equation}
	D_B(\rho_1,\rho_2) = \sqrt{2[1-F(\rho_1,\rho_2)]},
\end{equation}
where $F(\rho_1,\rho_2) = \Tr[(\sqrt{\rho_1}\rho_2\sqrt{\rho_1})^{1/2}]$ is the quantum fidelity between the states $\rho_1$ and $\rho_2$ \cite{Nielsen2010}. By evaluating the infinitesimal Bures distance explicitly, one finds that the Bures metric is indeed proportional to the QFI \cite{Sommers2003}.

In order to quantify the performance of an estimator and so the estimability of a certain parameter, 
a relevant figure of merit is the signal-to-noise ratio (SNR)
\begin{equation}
R_\beta \equiv \beta^2 F(\beta) \geq \frac{{\beta}^2}{\hbox{Var}(\beta)}
\label{9}
\end{equation}
which is larger for a better estimator.
We can easily derive an upper bound for this ratio using the Cram\'er-Rao inequality, obtaining
\begin{equation}
R_\beta\leq Q_\beta\equiv{\beta}^2H(\beta)
\label{10}
\end{equation}
which we refer to as the quantum signal-to-noise ratio (QSNR). The larger the quantities $R(\beta)$ and $Q(\beta)$ the smaller the relative error in the estimation of the parameter $\beta$.

\section{Quantum limits to precision in probing deformed commutators}
\label{sec:qfi_harmonic_oscillator}
We investigate the value of the QFI and the performance of a momentum measurement through the calculation of the FI as functions of $\beta$ for different states of the harmonic oscillator. In this way we find the estimability and the precision available through a momentum measurement as a function of the value of $\beta$, clarifying what values of $\beta$ could allow better estimation through experiments. 
In the following, we take $\hbar = 1$ and $k_B = 1$.
The parameters characterizing the harmonic oscillator, i.e. its mass $m$ and its pulsation $\omega$ are initially taken equal to $1$. We discuss the dependence of the QFI and FI on these parameters in Section \ref{sub:dependence_on_parameters}.

In the last section we discussed the tools of QET. In the problem at hand, however, standard QET has proven 
to be inaccurate, due to the particular geometry of the Hilbert space induced by the deformed commutators, Eq. \eqref{eq:modified_commutator}.
Indeed the scalar product has a non-trivial measure $\mu_\beta(p)$, Eq. \eqref{eq:integration_measure},  that 
depends on the parameter $\beta$. This in turn introduces a $\beta$-dependent measure in the sample space on 
which the probability $P(p|\beta)$ is defined, thus making the Cramer-Rao surpassable. 
This situation has been addressed recently in \cite{Seveso2016}, where an additional contribution 
to the FI is introduced. Let us redefine the FI as 
 \begin{equation}\label{eq:amended_fi}
	\FI(\beta) = F(\beta) + \mathcal{I}_\mu(\beta)\,,
\end{equation}
where
\begin{equation}\label{eq:measure_part_fi}
  \mathcal{I}_\mu(\beta)=\int dp\, \mu_\beta(p)\,P(p|\beta)\,[\partial_\beta \log \mu_\beta(p)]^2\;.
\end{equation}
Correspondingly, we redefine the SNR $\mathcal{R}(\beta) \equiv \beta^2 \FI(\beta)$.  
Being $\mathcal{I}_\mu$ a positive quantity, it follows that \eqref{eq:QCR} does not give the ultimate 
bound to the variance of any estimator of $\beta$.
It is not known whether $\FI$ in Eq. \eqref{eq:amended_fi} can be optimized over all possible 
quantum measurements so that a new quantum Cram\'er-Rao bound can be found.

\subsection{Pure states}
We first consider the estimation of $\beta$ from a measurement on the harmonic oscillator prepared in a pure state $\ket{\psi_\beta}$. Eq. \eqref{eq:qfi_pure_states_standard}, derived in Section \ref{sec:qet}, does not hold here because $\partial_\beta(\rho_\beta^2) \neq \partial_\beta \rho_\beta \rho_\beta + \rho_\beta \partial_\beta \rho_\beta$. Nevertheless, we can obtain a simplified expression for the QFI starting from Eq. \eqref{eq:qfi_explicit}.
We write $\rho_\beta = \sum_n p_n \ket{\phi_n}{\bra{\phi_n}}$, where $\ket{\phi_0} \equiv \ket{\psi_\beta}$, $p_n = \delta_{n0}$ and $\{\ket{\phi_n}\}_{n\neq 0}$ form a basis of the subspace orthogonal to $\ket{\psi_\beta}$. We obtain:
	\begin{align}
		H(\beta) & = 2 \sum_{\underset{\delta_{n0}+\delta_{m0} \neq 0}{n,m}} \frac{|\delta_{m0}\braket{\partial_\beta\phi_0|\phi_n}+\delta_{n0}\braket{\phi_n|\partial_\beta\phi_0}|^2}{\delta_{n0}+\delta_{m0}} \notag \\
		& = |\braket{\partial_\beta\phi_0|\phi_0}+\braket{\phi_0|\partial_\beta\phi_0}|^2+ 4 \sum_{n=1}^\infty |\braket{\phi_n|\partial_\beta\phi_0}|^2 \notag \\
		& = 4 \braket{\partial_\beta \psi_\beta | \partial_\beta \psi_\beta} - 4 \Im(\braket{\psi_\beta|\partial_\beta\psi_\beta})^2. \label{eq:qfi_pure_state}
	\end{align}

	Consider now a momentum measurement on the state described by the wavefunction $\psi_\beta(p)$. The probability of getting $p$ as an outcome is given by $P(p|\beta) = |\psi_\beta(p)|^2$, so the corresponding FI, Eq. \eqref{eq:amended_fi}, is
	\begin{equation}\label{eq:fi_momentum}
		\FI(\beta) = \int dp \left\{{\mu_\beta\frac{\left[\partial_\beta|\psi_\beta|^2\right]^2}{|\psi_\beta|^2} + |\psi_\beta|^2 \frac{[\partial_\beta \mu_\beta]^2}{\mu_\beta}}\right\}.
	\end{equation}
	
	Notice that if the wavefunction $\psi_\beta(p)$ is real, the first term of Eq. \eqref{eq:fi_momentum}, corresponding to $F(\beta)$, is equal to the QFI, Eq. \eqref{eq:qfi_pure_state}. Thus the FI for the momentum measurement is greater than the QFI and the standard Cram\'er-Rao bound is violated.
	
Using Eq. \eqref{eq:qfi_pure_state} and performing numerical integration of the scalar product, we calculate the QFI $H(\beta)$ for the first eigenstates of the harmonic oscillator. In all cases $H(\beta)$ is a decreasing function of $\beta$, but looking at the estimability $Q(\beta)$, which is the relevant quantity to consider, we have an increasing function of the parameter. 
If we consider eigenstates of higher energy, the QFI increases as can be checked numerically.

Since the value of $\beta$ is believed to be much smaller than one, the wavefunctions in Eqs. \eqref{eq:kempf} and \eqref{eq:chang} and the QFI, Eq. \eqref{eq:qfi_pure_state}, can be expanded around $\beta = 0$ in order to get analytic solutions which confirm the consistency of the numerical integrations. We obtain the following polynomial expressions:
\begin{align}
		H_{\psi_0}(\beta)&= \frac{9}{8} -\frac{53}{8} \beta + \frac{803}{32}\beta ^2+O\left(\beta^3\right) \label{eq:qfi_psi0} \\
H_{\psi_1}(\beta) & =\frac{45}{8} - \frac{351}8 \beta + \frac{7633}{32}\beta^2 +O\left(\beta^3\right) \label{eq:qfi_psi1} \\
H_{\psi_2}(\beta) & = \frac{123}{8}-\frac{1255}{8}\beta+\frac{36401 }{32}\beta^2+O\left(\beta^3\right) \label{eq:qfi_psi2}.
\end{align}

	\begin{figure}[t]
	\includegraphics{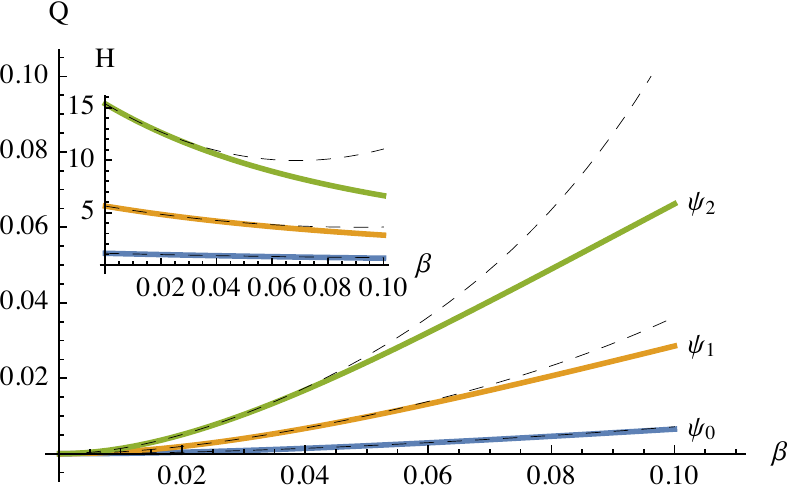}  
	\caption{(Color online) From bottom to top, estimability $Q(\beta)$  for the ground state (blue) and the first (orange) and second excited state (green) as obtained by numerical integration of the scalar product in Eq. \eqref{eq:qfi_pure_state}. In the inset, the QFI $H(\beta)$ for the same states. The dashed lines are obtained from the Taylor expansions of the QFI, Eqs. \eqref{eq:qfi_psi0} to \eqref{eq:qfi_psi2}. The estimability increases with by employing more excited states. The QFI decreases with $\beta$. In the region $\beta \lesssim 0.01$ the Taylor expansion provides a very good approximation.}
	\label{fig:eigenstates_qfi_fi}
	\end{figure}

Figure \ref{fig:eigenstates_qfi_fi} compares the analytical results with the numerical findings at various values of $\beta$. For $\beta \lesssim 0.01$, i.e. the expected range of values for $\beta$ \cite{Pikovski2012}, the approximation is very good with a relative error of at most $10^{-3}$.

The term $\mathcal{I}_\mu(\beta)$, for small $\beta$, reads
\begin{align}
	\mathcal{I}_{\mu,\psi_0}(\beta) & = \frac 34 - 3 \beta + 9 \beta^2 + O(\beta^3) \\
	\mathcal{I}_{\mu,\psi_1}(\beta) & = \frac{15}{4} - \frac{45 }{2}\beta  + \frac{405}{4} \beta ^2 + O(\beta^3) \\
	\mathcal{I}_{\mu,\psi_2}(\beta) & = \frac{39}{4} - \frac{165  }{2} \beta + \frac{2043 }{4} \beta ^2 + O(\beta^3)
\end{align}

Notice that $\mathcal{I}_{\mu,\psi_n}(\beta) \simeq 2/3 \, H_{\psi_n}(\beta) = F_{\psi_n}(\beta)$: the integration-measure term of $\FI(\beta)$ gives a relevant contribution to the estimability of $\beta$ through a momentum measurement.

\begin{figure}
\centering
\includegraphics[width=.49\columnwidth]{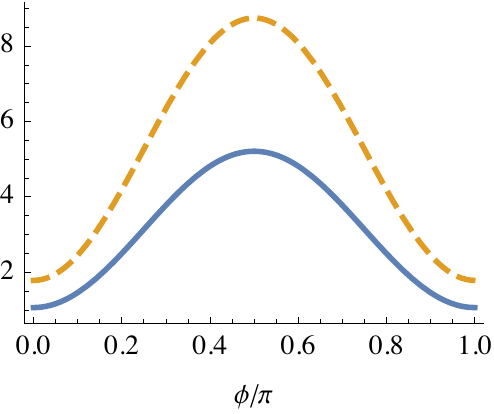}\includegraphics[width=.49\columnwidth]{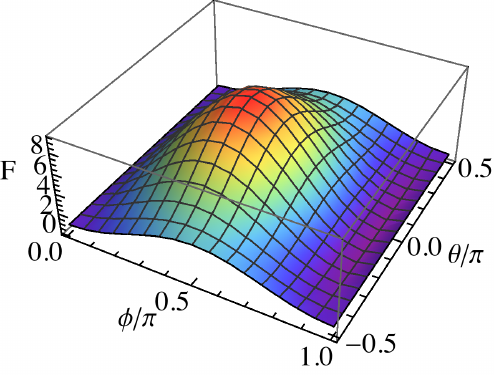}
\caption{(Color online) Left: QFI (solid blue) and FI (dashed orange) relative to the state $ |\psi\rangle = \cos(\phi)|\psi_0\rangle + \sin(\phi)|\psi_1\rangle$ as functions of $\phi$, with $\beta = 0.01$. The maximal values are reached when $\phi\rightarrow \pi/2$: among the superpositions of $\ket{\psi_0}$ and $\ket{\psi_1}$ the optimal state is the first excited state.
Right: FI as a function of the angles $\theta$ and $\phi$ for a superposition of the first three eigenstates, cf. Eq. \eqref{eq:qutrit_superposition}, for $\beta = 10^{-2}$. We can see that the maximal QFI is attained when $\theta = 0$ and $\phi \simeq 0.43 \pi$, i.e. when the system is in a superposition of the states $\ket{\psi_0}$ and $\ket{\psi_2}$.} \label{fig:superposition}
\end{figure}

We also studied the behavior of the QFI of the generic superposition of the ground and first excited state, to determine if the best estimability is attained by choosing the first excited state.
The system is thus described by
\begin{equation}\label{eq:qubit_superposition}
\ket{\psi} = \cos(\phi)\ket{\psi_0} + \sin(\phi)\ket{\psi_1}
\end{equation}
and the QFI is a function of the parameters $\beta$ and $\phi$.
The QFI has been calculated through numerical integration and it is shown in Fig. \ref{fig:superposition} (left): the maximal values of the function are obtained for $\phi\rightarrow \pi/2$ and $\phi\rightarrow 3/2 \;\pi$, i.e. the first excited state is the optimal state among those of Eq. \eqref{eq:qubit_superposition}. This can be seen numerically for arbitrary $\beta$ and analytically for small $\beta$, when the following expression holds:
\begin{equation}
	H(\beta) = H_{\psi_0}(\beta) + [H_{\psi_1}(\beta) - H_{\psi_0}(\beta)]\sin^2 \phi.
\end{equation}

We now consider a superposition of the first three eigenstates of the harmonic oscillator:
\begin{equation}\label{eq:qutrit_superposition}
	\ket{\psi} =  \cos \phi \ket{\psi_0} + \sin \phi \sin \theta \ket{\psi_1} + \sin \phi \cos \theta \ket{\psi_2}.
\end{equation}
In this case, the optimal state is not $\ket{\psi_2}$ as one would expect, given the previous result. The right panel of Fig. \ref{fig:superposition} shows the QFI for the superposition of the form of Eq. \eqref{eq:qutrit_superposition} as a function of $\theta$ and $\phi$. $\ket{\psi_2}$ is given by $\theta = 0$ and $\phi = \pi/2$ but the maximum is for $\theta = 0$ and $\phi \simeq 0.43 \pi$. Thus, in general, the eigentstates of the harmonic oscillator are not the states that give the best estimability.     

\subsection{Mixed states} 
\label{sub:mixed_states}
When the system is prepared in a mixed state $\rho_\beta = \sum_m p_m \ket{\psi_m}\bra{\psi_n}$, by expanding $\partial_\beta\rho$ in Eq. \eqref{eq:qfi_explicit} we obtain the following formula for the QFI:
\begin{equation}\label{eq:mixed_state_qfi}
\begin{split}
	H(\beta) =  2 \sum_{nm} &\frac{1}{p_n+p_m}| \partial_\beta p_m \delta_{mn} \\
	& + p_n \braket{\psi_m|\partial_\beta \psi_n} + p_m \braket{\partial_\beta \psi_m|\psi_n}|^2
\end{split}
\end{equation}

The FI $\FI(\beta)$ for the momentum measurement, Eq. \eqref{eq:amended_fi}, on the other hand, is given by the two contributions
\begin{equation}\label{eq:fi_mixed}
	F(\beta) = \sum_n p_n \int dp \mu_\beta(p) |\psi_n(p)|^2 \partial_\beta \ln |\psi_n(p)|^2
\end{equation}
and 
\begin{equation}
	\mathcal{I}_\mu = \sum_n p_n \int dp \mu_\beta(p) |\psi_n(p)|^2 \partial_\beta \ln \mu_\beta(p).
\end{equation}

As an example, we consider the estimation of $\beta$ from a measurement on the harmonic oscillator prepared in a generic statistical mixture of the ground and the first excited state.
The system is thus described by the statistical operator
\begin{equation}\label{eq:statistical_mixture}
\ket{\psi}\bra{\psi} = \cos(\theta)^2\ket{\psi_0}\bra{\psi_0} + \sin(\theta)^2\ket{\psi_1}\bra{\psi_1}.
\end{equation}

\begin{figure}
\includegraphics{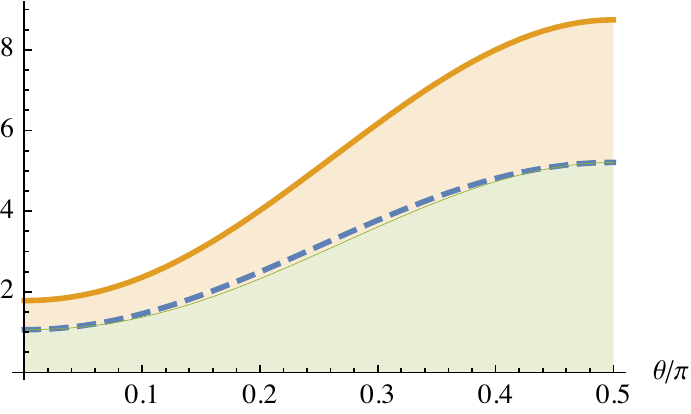}
\caption{(Color online) Comparison of QFI (dashed blue) and FI $\FI$ (solid orange) for the statistical mixture of the ground and first excited state, Eq. \eqref{eq:statistical_mixture}, as a function of $\theta$, with $\beta=0.01$. The two shaded regions represent the contributions to the FI coming from $F(\beta)$ (bottom, green) and $\mathcal{I}_\mu$ (top, orange), cf Eq. \eqref{eq:amended_fi}. The FI is much greater than the QFI due to the relevant contribution of the integration-measure term, $\mathcal{I}_\mu$. For $\theta = 0$ and $\theta =\Pi/2$, i.e. for pure states, $F(\beta)$ is equal to the QFI while for intermediate values of $\theta$ it is slightly slower, which means that the momentum measurement is not the optimal one (in the sense of the standard QET).}
\label{fig:qfi_mix_theta}
\end{figure}

We performed numerical integration of Eqs. \eqref{eq:mixed_state_qfi} and \eqref{eq:fi_mixed} and the results are shown in Fig. \ref{fig:qfi_mix_theta}.
The FI is much higher than the QFI due to the contribution of the term $\mathcal{I}_\mu$. While for $\theta\rightarrow 0 $ and $\theta\rightarrow \frac{\Pi}{2}$, i.e. when the state is pure, $F(\beta) = H(\beta)$, for intermediate values of $\theta$, $F(\beta)$ does not saturate the QFI, as we see in Fig.~\ref{fig:qfi_mix_theta}.
Thus, while in general the momentum measurement is not optimal for mixed states, the FI is much greater than the QFI due to the dependence of the geometry of the Hilbert space on $\beta$.
\subsection{Thermal state}
In a typical experimental setup it is generally challenging to prepare the oscillator in a pure state. Due to the interaction with the environment, the system will most likely be in a thermal state characterized by a temperature $T$. The density operator describing the state is then
\begin{equation}
	\rho_T =Z^{-1}  \sum_n e^{- E_n(\beta)/ T} \ket{\psi_n}\bra{\psi_n},
\end{equation}
where $Z = \sum_n e^{- E_n(\beta)/ T} $ is the partition function of the thermal distribution.
What is the maximum precision achievable if the oscillator is in the thermal state $\rho_T$? We focus on states with temperatures close to zero (compared to the ground state energy) so that only the lower eigenstates have significant populations. Indeed, the scalar products of the form $\braket{\partial_\beta \psi_n|\psi_m}$ that appear in Eq. \eqref{eq:mixed_state_qfi}, for high $m$ and $n$, involve highly oscillating functions and are thus hard to compute numerically to an acceptable accuracy.

\begin{figure}[t]  
\includegraphics{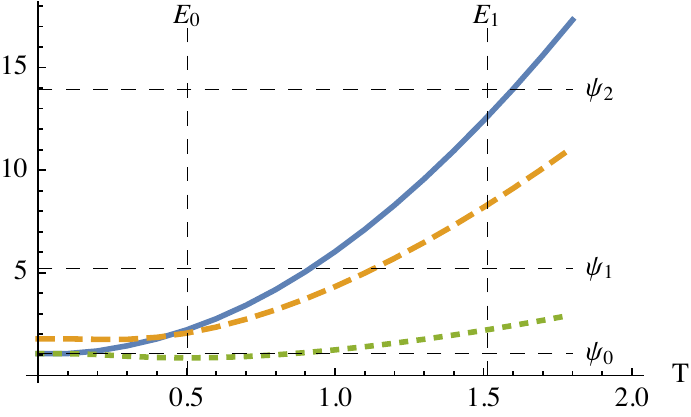}  
\caption{QFI $H(\beta)$ (solid blue), FI $\FI(\beta)$ (dashed orange) and $F(\beta)$ (dotted green) as functions of $T$ for $\beta = 0.01$ (with $\hbar = k_B = 1)$. The FI and QFI increase with temperature, because higher eigenstates of the oscillator are populated, but the performance of the momentum measurement gets worse as temperature increases. For $T$ close to zero $\FI(\beta)$ violates the CR bound, but at a temperature comparable with $E_0(\beta) \simeq \frac 12 + \frac \beta 4$, $\FI(\beta)$ gets lower than the QFI.}
\label{fig:thermal_state}
\end{figure}

As can be seen in Fig. \ref{fig:thermal_state}, QFI and FI are increasing functions of $T$. This is due to the fact that the population of higher eigenstates increases with $T$ and the QFI and FI increase with the energy of the eigenstate. When $T \lesssim E_0$, $\FI(\beta)$ is greater than $H(\beta)$, violating the quantum Cramér-Rao bound; on the other hand, when the temperature increases, the momentum measurement is not optimal anymore

\subsection{Dependence on $m$ and $\omega$}
\label{sub:dependence_on_parameters}
\begin{figure}[t]
\includegraphics{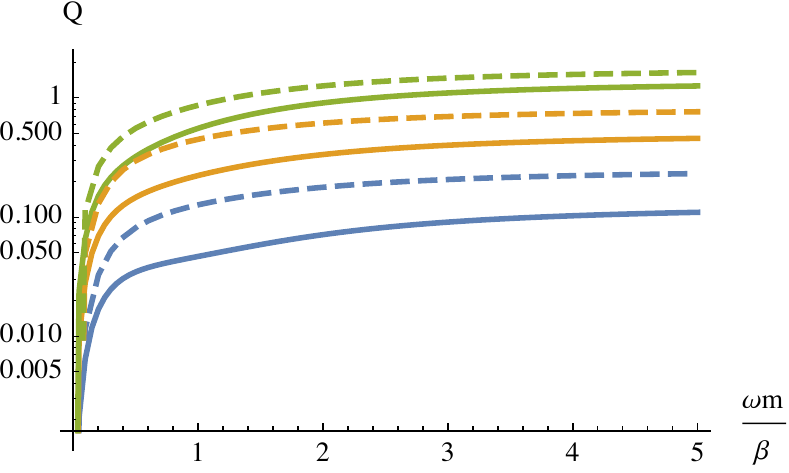}  
\caption{(Color online) Log-plot of the quantum estimability $Q(\beta)$ (solid lines) and estimability $\mathcal{R}(\beta)$ for the momentum measurement (dashed lines) as functions of $\omega m /\beta$ for the pure states (from bottom to top) $\psi_0$, $\psi_1$ and $\psi_2$. The plots do not depend on $\beta$. $Q(\beta)$  and $\mathcal{R}(\beta)$ increase with the product $\omega m$ and reach the limits reported in Eq. \eqref{eq:limits_large_omega_m}.} \label{fig:omega_m}
\end{figure}

In the previous Section we have shown the behavior of the QFI as a function of $\beta$ assuming $\omega = 1$ and $m = 1$. In this Section we show how the QFI depends on the mass and frequency of the harmonic oscillator.

By looking at Eqs. \eqref{eq:kempf} and \eqref{eq:chang}, we notice that the eigenstates of the harmonic oscillator depend on $m$ and $\omega$ only through the product $\omega m \beta$ in the term $\lambda$. 

 As we see in Fig. \ref{fig:omega_m}, in the example of the ground state, $H(\beta)$ is an increasing function of $\omega m$. We can obtain analytically the limits for $\omega m \rightarrow 0$
\begin{equation}\label{eq:limit_small_omega_m}
	H(\beta) \underset{\omega m \rightarrow 0}{\longrightarrow} 0
\end{equation}
and $\omega m \gg \beta$:
\begin{equation}\label{eq:limits_large_omega_m}
	Q_{\psi_0}(\beta) {\sim} \frac 1 {8 }, \qquad
	Q_{\psi_1}(\beta) {\sim} \frac 1 {2 }, \qquad
	Q_{\psi_2}(\beta) {\sim} \frac {11} {8}.
\end{equation}
As for the FI, we find that, for large $\omega m$, the SNR is twice the QSNR: $\mathcal{R}_{\psi_n}(\beta) \sim 2 Q_{\psi_n}(\beta)$. 
Equation \eqref{eq:limits_large_omega_m} shows that the SNR and QSNR of $\beta$ do not depend on it's value for large enough $\omega m$. 

\section{Conclusions} 
\label{sec:conclusions}
Although a minimal length at the Planck scale is predicted by many theories of quantum gravity, the lack of theoretical 
predictions about its value and the formidable technological challenges required, experimental tests have been so far inconclusive. The aim of this paper is to provide theoretical tools to asses the best achievable precision in the estimation of the deformation of the canonical commutation relations induced by the minimal length. We focused on measurements on a
harmonic oscillator, a relevant testbed both from a theoretical point of view, as it is analytically solvable, and from an experimental point of view, since experiments can and have been made with nano-mechanical and opto-mechanical oscillators. 

We have shown that a measurement of the momentum is optimal if the oscillator is in a pure state and the achievable precision goes beyond the bounds of standard quantum estimation theory. This is a relevant result, due to the altered geometry of the Hilbert space, and shows the necessity of redefining the quantities of QET in a more general way \cite{Seveso2016}.

Our results indicate that the estimability improves by preparing the oscillator in a higher energy eigenstate. Moreover, increasing the mass and frequency of the oscillator allows for better precision and the temperature is not detrimental for the probing, although the momentum measurement ceases to be the optimal measurement as the temperature increases above the energy of the ground state.

\begin{acknowledgments}
	MR thanks Francesco Albarelli, Nicola Seveso for fruitful discussions and the user Martin Nicholson for a useful discussion on \url{math.stackexchange.com}. This work has been supported by EU through the collaborative Project QuProCS (Grant Agreement 641277) and by Università degli Studi di Milano through the H2020 Transition Grant 15-6-3008000-625.
\end{acknowledgments}

\appendix
\section{Relation between the solutions of the harmonic oscillator in the momentum basis}
\label{app:solutions}
Here we show the relation between the two solutions of
the harmonic oscillator. We also find the normalization
constant $\mathcal{N}_n$ for the solution \eqref{eq:kempf}, involving the hypergeometric function.

The solution of \cite{Chang2002} is normalized.
Let us start from Eq. \eqref{eq:kempf} and show that it can be cast
to the form of Eq. \eqref{eq:chang}. We assume that $n$ is even, i.e.
we set $n = 2\nu$, with $\nu\in N $. The case with $n$ odd is
analogous.
The argument of $\hypgf$ in Eq. \eqref{eq:kempf} is complex, but we
can apply Kummers' quadratic transformation (15.8.18)
from \cite{NistHB} to obtain
\begin{equation}
\begin{split}
\psi_n(p) = {}& \mathcal{N}_n(1+\beta^2)^{-\frac{\lambda}{2}-\nu} \\ & \hypgf\left(-\nu,\frac{1}{2}-\lambda-\nu;1-\lambda-2\nu;1+\beta p^2\right).
\end{split}
\label{23}
\end{equation}
Next we apply Eq. (15.8.6) of \cite{NistHB} to invert the argument
of $\hypgf$: we end up with
\begin{align}
\psi_n(p) = {} & \frac{\mathcal{N}_n\sqrt{\pi}(-1)^{\nu}4^{-\lambda-\nu}}{(1+\beta p^2)^{\frac{\lambda}{2}}} \frac{\sec(\pi\lambda)\Gamma(-\lambda-2\nu+1)}{\Gamma(\lambda+\frac{1}{2})+\Gamma(-2\lambda-2\nu+1)} \notag \\ 
&\hypgf\left(-\nu,\lambda+\nu;\lambda+\frac{1}{2};\frac{1}{1+\beta p^2}\right).
\label{24}
\end{align}
By applying Eq. (20) of \cite{MathWorld} and by plugging back $n$, we finally reach the functional form of Eq. \eqref{eq:chang}:
\begin{equation}
\begin{split}
\psi_n(p)  ={} & \mathcal{N}_n\frac{i^n n! \sin(\pi\lambda)\Gamma(\lambda)\Gamma(1-n-\lambda)}{(-2)^n\pi}\\ &(1+\beta p^2)^{-\frac{\lambda}{2}}C_n^{(\lambda)}\left(\sqrt{\frac{\beta p^2}{1+\beta p^2}}\right).
\end{split}
\label{eq:last_passage}
\end{equation}
The same result can be obtained for odd $n$ by applying Eq. (21) of \cite{MathWorld}. 

By comparing Eq. \eqref{eq:last_passage} and \eqref{eq:kempf} we obtain
an expression for the normalization constant
\begin{equation}\label{eq:normalization_constant}
\mathcal{N}_n=\frac{(-i)^n\sqrt{\pi}\sqrt[4]{\beta}2^{\lambda+n-\frac{1}{2}}}{\sin(\pi\lambda)\Gamma(1-n-\lambda)}\sqrt{\frac{\lambda+n}{n!\Gamma(n+2\lambda)}}.
\end{equation}
 
\bibliography{betagup.bib}

\begin{thebibliography}{34}%
\makeatletter
\providecommand \@ifxundefined [1]{%
 \@ifx{#1\undefined}
}%
\providecommand \@ifnum [1]{%
 \ifnum #1\expandafter \@firstoftwo
 \else \expandafter \@secondoftwo
 \fi
}%
\providecommand \@ifx [1]{%
 \ifx #1\expandafter \@firstoftwo
 \else \expandafter \@secondoftwo
 \fi
}%
\providecommand \natexlab [1]{#1}%
\providecommand \enquote  [1]{``#1''}%
\providecommand \bibnamefont  [1]{#1}%
\providecommand \bibfnamefont [1]{#1}%
\providecommand \citenamefont [1]{#1}%
\providecommand \href@noop [0]{\@secondoftwo}%
\providecommand \href [0]{\begingroup \@sanitize@url \@href}%
\providecommand \@href[1]{\@@startlink{#1}\@@href}%
\providecommand \@@href[1]{\endgroup#1\@@endlink}%
\providecommand \@sanitize@url [0]{\catcode `\\12\catcode `\$12\catcode
  `\&12\catcode `\#12\catcode `\^12\catcode `\_12\catcode `\%12\relax}%
\providecommand \@@startlink[1]{}%
\providecommand \@@endlink[0]{}%
\providecommand \url  [0]{\begingroup\@sanitize@url \@url }%
\providecommand \@url [1]{\endgroup\@href {#1}{\urlprefix }}%
\providecommand \urlprefix  [0]{URL }%
\providecommand \Eprint [0]{\href }%
\providecommand \doibase [0]{http://dx.doi.org/}%
\providecommand \selectlanguage [0]{\@gobble}%
\providecommand \bibinfo  [0]{\@secondoftwo}%
\providecommand \bibfield  [0]{\@secondoftwo}%
\providecommand \translation [1]{[#1]}%
\providecommand \BibitemOpen [0]{}%
\providecommand \bibitemStop [0]{}%
\providecommand \bibitemNoStop [0]{.\EOS\space}%
\providecommand \EOS [0]{\spacefactor3000\relax}%
\providecommand \BibitemShut  [1]{\csname bibitem#1\endcsname}%
\let\auto@bib@innerbib\@empty
\bibitem [{\citenamefont {Garay}(1995)}]{Garay1995}%
  \BibitemOpen
  \bibfield  {author} {\bibinfo {author} {\bibfnamefont {L.~J.}\ \bibnamefont
  {Garay}},\ }\href {\doibase 10.1142/S0217751X95000085} {\bibfield  {journal}
  {\bibinfo  {journal} {Int. J. Mod. Phys. A}\ }\textbf {\bibinfo {volume}
  {10}},\ \bibinfo {pages} {145} (\bibinfo {year} {1995})}\BibitemShut
  {NoStop}%
\bibitem [{\citenamefont {Hossenfelder}(2013)}]{Hossenfelder2013}%
  \BibitemOpen
  \bibfield  {author} {\bibinfo {author} {\bibfnamefont {S.}~\bibnamefont
  {Hossenfelder}},\ }\href {\doibase 10.12942/lrr-2013-2} {\bibfield  {journal}
  {\bibinfo  {journal} {Living Rev. Relativity}\ }\textbf {\bibinfo {volume}
  {16}},\ \bibinfo {pages} {2} (\bibinfo {year} {2013})}\BibitemShut {NoStop}%
\bibitem [{\citenamefont {Maggiore}(1993)}]{Maggiore1993}%
  \BibitemOpen
  \bibfield  {author} {\bibinfo {author} {\bibfnamefont {M.}~\bibnamefont
  {Maggiore}},\ }\href {\doibase 10.1016/0370-2693(93)90785-g} {\bibfield
  {journal} {\bibinfo  {journal} {Phys. Lett. B}\ }\textbf {\bibinfo {volume}
  {319}},\ \bibinfo {pages} {83} (\bibinfo {year} {1993})}\BibitemShut
  {NoStop}%
\bibitem [{\citenamefont {Kempf}(1994)}]{Kempf1994}%
  \BibitemOpen
  \bibfield  {author} {\bibinfo {author} {\bibfnamefont {A.}~\bibnamefont
  {Kempf}},\ }\href {\doibase 10.1063/1.530798} {\bibfield  {journal} {\bibinfo
   {journal} {J. Math. Phys.}\ }\textbf {\bibinfo {volume} {35}},\ \bibinfo
  {pages} {4483} (\bibinfo {year} {1994})}\BibitemShut {NoStop}%
\bibitem [{\citenamefont {Kempf}\ \emph {et~al.}(1995)\citenamefont {Kempf},
  \citenamefont {Mangano},\ and\ \citenamefont {Mann}}]{Kempf1995}%
  \BibitemOpen
  \bibfield  {author} {\bibinfo {author} {\bibfnamefont {A.}~\bibnamefont
  {Kempf}}, \bibinfo {author} {\bibfnamefont {G.}~\bibnamefont {Mangano}}, \
  and\ \bibinfo {author} {\bibfnamefont {R.~B.}\ \bibnamefont {Mann}},\ }\href
  {\doibase 10.1103/PhysRevD.52.1108} {\bibfield  {journal} {\bibinfo
  {journal} {Phys. Rev. D}\ }\textbf {\bibinfo {volume} {52}},\ \bibinfo
  {pages} {1108} (\bibinfo {year} {1995})}\BibitemShut {NoStop}%
\bibitem [{\citenamefont {Kempf}(1997)}]{Kempf1997}%
  \BibitemOpen
  \bibfield  {author} {\bibinfo {author} {\bibfnamefont {A.}~\bibnamefont
  {Kempf}},\ }\href {\doibase 10.1088/0305-4470/30/6/030} {\bibfield  {journal}
  {\bibinfo  {journal} {J. Phys. A: Math. Gen.}\ }\textbf {\bibinfo {volume}
  {30}},\ \bibinfo {pages} {2093} (\bibinfo {year} {1997})}\BibitemShut
  {NoStop}%
\bibitem [{\citenamefont {Milburn}(2006)}]{Milburn2006}%
  \BibitemOpen
  \bibfield  {author} {\bibinfo {author} {\bibfnamefont {G.~J.}\ \bibnamefont
  {Milburn}},\ }\href {\doibase 10.1088/1367-2630/8/6/096} {\bibfield
  {journal} {\bibinfo  {journal} {New J. Phys.}\ }\textbf {\bibinfo {volume}
  {8}},\ \bibinfo {pages} {96} (\bibinfo {year} {2006})}\BibitemShut {NoStop}%
\bibitem [{\citenamefont {Lewis}\ and\ \citenamefont
  {Takeuchi}(2011)}]{Lewis2011}%
  \BibitemOpen
  \bibfield  {author} {\bibinfo {author} {\bibfnamefont {Z.}~\bibnamefont
  {Lewis}}\ and\ \bibinfo {author} {\bibfnamefont {T.}~\bibnamefont
  {Takeuchi}},\ }\href {\doibase 10.1103/PhysRevD.84.105029} {\bibfield
  {journal} {\bibinfo  {journal} {Phys. Rev. D}\ }\textbf {\bibinfo {volume}
  {84}},\ \bibinfo {pages} {105029} (\bibinfo {year} {2011})}\BibitemShut
  {NoStop}%
\bibitem [{\citenamefont {Chang}\ \emph {et~al.}(2002)\citenamefont {Chang},
  \citenamefont {Minic}, \citenamefont {Okamura},\ and\ \citenamefont
  {Takeuchi}}]{Chang2002}%
  \BibitemOpen
  \bibfield  {author} {\bibinfo {author} {\bibfnamefont {L.~N.}\ \bibnamefont
  {Chang}}, \bibinfo {author} {\bibfnamefont {D.}~\bibnamefont {Minic}},
  \bibinfo {author} {\bibfnamefont {N.}~\bibnamefont {Okamura}}, \ and\
  \bibinfo {author} {\bibfnamefont {T.}~\bibnamefont {Takeuchi}},\ }\href
  {\doibase 10.1103/PhysRevD.65.125027} {\bibfield  {journal} {\bibinfo
  {journal} {Phys. Rev. D}\ }\textbf {\bibinfo {volume} {65}},\ \bibinfo
  {pages} {125027} (\bibinfo {year} {2002})}\BibitemShut {NoStop}%
\bibitem [{\citenamefont {Das}\ and\ \citenamefont {Vagenas}(2008)}]{Das2008}%
  \BibitemOpen
  \bibfield  {author} {\bibinfo {author} {\bibfnamefont {S.}~\bibnamefont
  {Das}}\ and\ \bibinfo {author} {\bibfnamefont {E.~C.}\ \bibnamefont
  {Vagenas}},\ }\href {\doibase 10.1103/PhysRevLett.101.221301} {\bibfield
  {journal} {\bibinfo  {journal} {Phys. Rev. Lett.}\ }\textbf {\bibinfo
  {volume} {101}},\ \bibinfo {pages} {221301} (\bibinfo {year}
  {2008})}\BibitemShut {NoStop}%
\bibitem [{\citenamefont {Harbach}\ \emph {et~al.}(2004)\citenamefont
  {Harbach}, \citenamefont {Hossenfelder}, \citenamefont {Bleicher},\ and\
  \citenamefont {Stöcker}}]{Harbach2004}%
  \BibitemOpen
  \bibfield  {author} {\bibinfo {author} {\bibfnamefont {U.}~\bibnamefont
  {Harbach}}, \bibinfo {author} {\bibfnamefont {S.}~\bibnamefont
  {Hossenfelder}}, \bibinfo {author} {\bibfnamefont {M.}~\bibnamefont
  {Bleicher}}, \ and\ \bibinfo {author} {\bibfnamefont {H.}~\bibnamefont
  {Stöcker}},\ }\href {\doibase
  http://dx.doi.org/10.1016/j.physletb.2004.01.051} {\bibfield  {journal}
  {\bibinfo  {journal} {Phys. Lett. B}\ }\textbf {\bibinfo {volume} {584}},\
  \bibinfo {pages} {109 } (\bibinfo {year} {2004})}\BibitemShut {NoStop}%
\bibitem [{\citenamefont {Sprenger}\ \emph {et~al.}(2011)\citenamefont
  {Sprenger}, \citenamefont {Nicolini},\ and\ \citenamefont
  {Bleicher}}]{Sprenger2011}%
  \BibitemOpen
  \bibfield  {author} {\bibinfo {author} {\bibfnamefont {M.}~\bibnamefont
  {Sprenger}}, \bibinfo {author} {\bibfnamefont {P.}~\bibnamefont {Nicolini}},
  \ and\ \bibinfo {author} {\bibfnamefont {M.}~\bibnamefont {Bleicher}},\
  }\href {\doibase 10.1088/0264-9381/28/23/235019} {\bibfield  {journal}
  {\bibinfo  {journal} {Class. Quantum Grav.}\ }\textbf {\bibinfo {volume}
  {28}},\ \bibinfo {pages} {235019} (\bibinfo {year} {2011})}\BibitemShut
  {NoStop}%
\bibitem [{\citenamefont {Pikovski}\ \emph {et~al.}(2012)\citenamefont
  {Pikovski}, \citenamefont {Vanner}, \citenamefont {Aspelmeyer}, \citenamefont
  {Kim},\ and\ \citenamefont {Brukner}}]{Pikovski2012}%
  \BibitemOpen
  \bibfield  {author} {\bibinfo {author} {\bibfnamefont {I.}~\bibnamefont
  {Pikovski}}, \bibinfo {author} {\bibfnamefont {M.~R.}\ \bibnamefont
  {Vanner}}, \bibinfo {author} {\bibfnamefont {M.}~\bibnamefont {Aspelmeyer}},
  \bibinfo {author} {\bibfnamefont {M.~S.}\ \bibnamefont {Kim}}, \ and\
  \bibinfo {author} {\bibfnamefont {{\v{C}}.}~\bibnamefont {Brukner}},\ }\href
  {\doibase 10.1038/nphys2262} {\bibfield  {journal} {\bibinfo  {journal}
  {Nature Phys.}\ }\textbf {\bibinfo {volume} {8}},\ \bibinfo {pages} {393}
  (\bibinfo {year} {2012})}\BibitemShut {NoStop}%
\bibitem [{\citenamefont {Bawaj}\ \emph {et~al.}(2015)\citenamefont {Bawaj},
  \citenamefont {Biancofiore}, \citenamefont {Bonaldi}, \citenamefont
  {Bonfigli}, \citenamefont {Borrielli}, \citenamefont {{Di Giuseppe}},
  \citenamefont {Marconi}, \citenamefont {Marino}, \citenamefont {Natali},
  \citenamefont {Pontin}, \citenamefont {Prodi}, \citenamefont {Serra},
  \citenamefont {Vitali},\ and\ \citenamefont {Marin}}]{Bawaj2015}%
  \BibitemOpen
  \bibfield  {author} {\bibinfo {author} {\bibfnamefont {M.}~\bibnamefont
  {Bawaj}}, \bibinfo {author} {\bibfnamefont {C.}~\bibnamefont {Biancofiore}},
  \bibinfo {author} {\bibfnamefont {M.}~\bibnamefont {Bonaldi}}, \bibinfo
  {author} {\bibfnamefont {F.}~\bibnamefont {Bonfigli}}, \bibinfo {author}
  {\bibfnamefont {A.}~\bibnamefont {Borrielli}}, \bibinfo {author}
  {\bibfnamefont {G.}~\bibnamefont {{Di Giuseppe}}}, \bibinfo {author}
  {\bibfnamefont {L.}~\bibnamefont {Marconi}}, \bibinfo {author} {\bibfnamefont
  {F.}~\bibnamefont {Marino}}, \bibinfo {author} {\bibfnamefont
  {R.}~\bibnamefont {Natali}}, \bibinfo {author} {\bibfnamefont
  {A.}~\bibnamefont {Pontin}}, \bibinfo {author} {\bibfnamefont {G.~A.}\
  \bibnamefont {Prodi}}, \bibinfo {author} {\bibfnamefont {E.}~\bibnamefont
  {Serra}}, \bibinfo {author} {\bibfnamefont {D.}~\bibnamefont {Vitali}}, \
  and\ \bibinfo {author} {\bibfnamefont {F.}~\bibnamefont {Marin}},\ }\href
  {\doibase 10.1038/ncomms8503} {\bibfield  {journal} {\bibinfo  {journal}
  {Nat. Commun.}\ }\textbf {\bibinfo {volume} {6}},\ \bibinfo {pages} {7503}
  (\bibinfo {year} {2015})}\BibitemShut {NoStop}%
\bibitem [{\citenamefont {Helstrom}(1976)}]{Helstrom1976}%
  \BibitemOpen
  \bibfield  {author} {\bibinfo {author} {\bibfnamefont {C.~W.}\ \bibnamefont
  {Helstrom}},\ }\href@noop {} {\emph {\bibinfo {title} {{Quantum Detection and
  Estimation Theory}}}}\ (\bibinfo  {publisher} {Academic Press},\ \bibinfo
  {address} {New York},\ \bibinfo {year} {1976})\BibitemShut {NoStop}%
\bibitem [{\citenamefont {Holevo}(2001)}]{Holevo2001}%
  \BibitemOpen
  \bibfield  {author} {\bibinfo {author} {\bibfnamefont {A.~S.}\ \bibnamefont
  {Holevo}},\ }\href {https://books.google.co.uk/books?id=uGl188JPxdQC} {\emph
  {\bibinfo {title} {{Statistical Structure of Quantum Theory}}}},\ Lecture
  Notes in Physics Monographs\ (\bibinfo  {publisher} {Springer},\ \bibinfo
  {year} {2001})\BibitemShut {NoStop}%
\bibitem [{\citenamefont {Braunstein}\ and\ \citenamefont
  {Caves}(1994)}]{Braunstein1994}%
  \BibitemOpen
  \bibfield  {author} {\bibinfo {author} {\bibfnamefont {S.~L.}\ \bibnamefont
  {Braunstein}}\ and\ \bibinfo {author} {\bibfnamefont {C.~M.}\ \bibnamefont
  {Caves}},\ }\href {\doibase 10.1103/PhysRevLett.72.3439} {\bibfield
  {journal} {\bibinfo  {journal} {Phys. Rev. Lett.}\ }\textbf {\bibinfo
  {volume} {72}},\ \bibinfo {pages} {3439} (\bibinfo {year}
  {1994})}\BibitemShut {NoStop}%
\bibitem [{\citenamefont {Paris}(2009)}]{Paris2009}%
  \BibitemOpen
  \bibfield  {author} {\bibinfo {author} {\bibfnamefont {M.~G.~A.}\
  \bibnamefont {Paris}},\ }\href {\doibase 10.1142/S0219749909004839}
  {\bibfield  {journal} {\bibinfo  {journal} {Int. J. Quantum Inf.}\ }\textbf
  {\bibinfo {volume} {7}},\ \bibinfo {pages} {125} (\bibinfo {year}
  {2009})}\BibitemShut {NoStop}%
\bibitem [{\citenamefont {Cram{\'{e}}r}(1946)}]{Cramer1946}%
  \BibitemOpen
  \bibfield  {author} {\bibinfo {author} {\bibfnamefont {H.}~\bibnamefont
  {Cram{\'{e}}r}},\ }\href@noop {} {\emph {\bibinfo {title} {{Mathematical
  Methods of Statistics}}}}\ (\bibinfo  {publisher} {Princeton Univ. Press},\
  \bibinfo {address} {Princeton},\ \bibinfo {year} {1946})\BibitemShut
  {NoStop}%
\bibitem [{\citenamefont {Giovannetti}(2004)}]{Giovannetti2004}%
  \BibitemOpen
  \bibfield  {author} {\bibinfo {author} {\bibfnamefont {V.}~\bibnamefont
  {Giovannetti}},\ }\href {\doibase 10.1126/science.1104149} {\bibfield
  {journal} {\bibinfo  {journal} {Science}\ }\textbf {\bibinfo {volume}
  {306}},\ \bibinfo {pages} {1330} (\bibinfo {year} {2004})}\BibitemShut
  {NoStop}%
\bibitem [{\citenamefont {Giovannetti}\ \emph {et~al.}(2011)\citenamefont
  {Giovannetti}, \citenamefont {Lloyd},\ and\ \citenamefont
  {Maccone}}]{Giovannetti2011}%
  \BibitemOpen
  \bibfield  {author} {\bibinfo {author} {\bibfnamefont {V.}~\bibnamefont
  {Giovannetti}}, \bibinfo {author} {\bibfnamefont {S.}~\bibnamefont {Lloyd}},
  \ and\ \bibinfo {author} {\bibfnamefont {L.}~\bibnamefont {Maccone}},\ }\href
  {\doibase 10.1038/nphoton.2011.35} {\bibfield  {journal} {\bibinfo  {journal}
  {Nature Photon.}\ }\textbf {\bibinfo {volume} {5}},\ \bibinfo {pages} {222}
  (\bibinfo {year} {2011})}\BibitemShut {NoStop}%
\bibitem [{\citenamefont {Monras}\ and\ \citenamefont
  {Paris}(2007)}]{Monras2007}%
  \BibitemOpen
  \bibfield  {author} {\bibinfo {author} {\bibfnamefont {A.}~\bibnamefont
  {Monras}}\ and\ \bibinfo {author} {\bibfnamefont {M.~G.~A.}\ \bibnamefont
  {Paris}},\ }\href {\doibase 10.1103/PhysRevLett.98.160401} {\bibfield
  {journal} {\bibinfo  {journal} {Phys. Rev. Lett.}\ }\textbf {\bibinfo
  {volume} {98}},\ \bibinfo {pages} {160401} (\bibinfo {year}
  {2007})}\BibitemShut {NoStop}%
\bibitem [{\citenamefont {Nagata}\ \emph {et~al.}(2007)\citenamefont {Nagata},
  \citenamefont {Okamoto}, \citenamefont {O'Brien}, \citenamefont {Sasaki},\
  and\ \citenamefont {Takeuchi}}]{Nagata2007}%
  \BibitemOpen
  \bibfield  {author} {\bibinfo {author} {\bibfnamefont {T.}~\bibnamefont
  {Nagata}}, \bibinfo {author} {\bibfnamefont {R.}~\bibnamefont {Okamoto}},
  \bibinfo {author} {\bibfnamefont {J.~L.}\ \bibnamefont {O'Brien}}, \bibinfo
  {author} {\bibfnamefont {K.}~\bibnamefont {Sasaki}}, \ and\ \bibinfo {author}
  {\bibfnamefont {S.}~\bibnamefont {Takeuchi}},\ }\href {\doibase
  10.1126/science.1138007} {\bibfield  {journal} {\bibinfo  {journal}
  {Science}\ }\textbf {\bibinfo {volume} {316}},\ \bibinfo {pages} {726}
  (\bibinfo {year} {2007})}\BibitemShut {NoStop}%
\bibitem [{\citenamefont {Berni}\ \emph {et~al.}(2015)\citenamefont {Berni},
  \citenamefont {Gehring}, \citenamefont {Nielsen}, \citenamefont
  {H\"{a}ndchen}, \citenamefont {Paris},\ and\ \citenamefont
  {Andersen}}]{Berni2015}%
  \BibitemOpen
  \bibfield  {author} {\bibinfo {author} {\bibfnamefont {A.~A.}\ \bibnamefont
  {Berni}}, \bibinfo {author} {\bibfnamefont {T.}~\bibnamefont {Gehring}},
  \bibinfo {author} {\bibfnamefont {B.~M.}\ \bibnamefont {Nielsen}}, \bibinfo
  {author} {\bibfnamefont {V.}~\bibnamefont {H\"{a}ndchen}}, \bibinfo {author}
  {\bibfnamefont {M.~G.~A.}\ \bibnamefont {Paris}}, \ and\ \bibinfo {author}
  {\bibfnamefont {U.~L.}\ \bibnamefont {Andersen}},\ }\href {\doibase
  10.1038/nphoton.2015.139} {\bibfield  {journal} {\bibinfo  {journal} {Nature
  Photon.}\ }\textbf {\bibinfo {volume} {9}},\ \bibinfo {pages} {577} (\bibinfo
  {year} {2015})}\BibitemShut {NoStop}%
\bibitem [{\citenamefont {Meyer}\ \emph {et~al.}(2001)\citenamefont {Meyer},
  \citenamefont {Rowe}, \citenamefont {Kielpinski}, \citenamefont {Sackett},
  \citenamefont {Itano}, \citenamefont {Monroe},\ and\ \citenamefont
  {Wineland}}]{Meyer2001}%
  \BibitemOpen
  \bibfield  {author} {\bibinfo {author} {\bibfnamefont {V.}~\bibnamefont
  {Meyer}}, \bibinfo {author} {\bibfnamefont {M.~A.}\ \bibnamefont {Rowe}},
  \bibinfo {author} {\bibfnamefont {D.}~\bibnamefont {Kielpinski}}, \bibinfo
  {author} {\bibfnamefont {C.~A.}\ \bibnamefont {Sackett}}, \bibinfo {author}
  {\bibfnamefont {W.~M.}\ \bibnamefont {Itano}}, \bibinfo {author}
  {\bibfnamefont {C.}~\bibnamefont {Monroe}}, \ and\ \bibinfo {author}
  {\bibfnamefont {D.~J.}\ \bibnamefont {Wineland}},\ }\href {\doibase
  10.1103/physrevlett.86.5870} {\bibfield  {journal} {\bibinfo  {journal}
  {Phys. Rev. Lett.}\ }\textbf {\bibinfo {volume} {86}},\ \bibinfo {pages}
  {5870} (\bibinfo {year} {2001})}\BibitemShut {NoStop}%
\bibitem [{\citenamefont {Leibfried}\ \emph {et~al.}(2004)\citenamefont
  {Leibfried}, \citenamefont {Barrett}, \citenamefont {Schaetz}, \citenamefont
  {Britton}, \citenamefont {Chiaverini}, \citenamefont {Itano}, \citenamefont
  {Jost}, \citenamefont {Langer},\ and\ \citenamefont
  {Wineland}}]{Leibfried2004}%
  \BibitemOpen
  \bibfield  {author} {\bibinfo {author} {\bibfnamefont {D.}~\bibnamefont
  {Leibfried}}, \bibinfo {author} {\bibfnamefont {M.~D.}\ \bibnamefont
  {Barrett}}, \bibinfo {author} {\bibfnamefont {T.}~\bibnamefont {Schaetz}},
  \bibinfo {author} {\bibfnamefont {J.}~\bibnamefont {Britton}}, \bibinfo
  {author} {\bibfnamefont {J.}~\bibnamefont {Chiaverini}}, \bibinfo {author}
  {\bibfnamefont {W.~M.}\ \bibnamefont {Itano}}, \bibinfo {author}
  {\bibfnamefont {J.~D.}\ \bibnamefont {Jost}}, \bibinfo {author}
  {\bibfnamefont {C.}~\bibnamefont {Langer}}, \ and\ \bibinfo {author}
  {\bibfnamefont {D.~J.}\ \bibnamefont {Wineland}},\ }\href {\doibase
  10.1126/science.1097576} {\bibfield  {journal} {\bibinfo  {journal}
  {Science}\ }\textbf {\bibinfo {volume} {304}},\ \bibinfo {pages} {1476}
  (\bibinfo {year} {2004})}\BibitemShut {NoStop}%
\bibitem [{\citenamefont {Seveso}\ \emph {et~al.}()\citenamefont {Seveso},
  \citenamefont {Rossi},\ and\ \citenamefont {Paris}}]{Seveso2016}%
  \BibitemOpen
  \bibfield  {author} {\bibinfo {author} {\bibfnamefont {L.}~\bibnamefont
  {Seveso}}, \bibinfo {author} {\bibfnamefont {M.~A.~C.}\ \bibnamefont
  {Rossi}}, \ and\ \bibinfo {author} {\bibfnamefont {M.~G.~A.}\ \bibnamefont
  {Paris}},\ }\href@noop {} {\enquote {\bibinfo {title} {{New Cram{\'e}r-Rao
  bounds for quantum metrology}},}\ }\Eprint
  {http://arxiv.org/abs/arXiv:1605.08653} {arXiv:1605.08653} \BibitemShut
  {NoStop}%
\bibitem [{\citenamefont {{Deriglazov, A. A.}}\ and\ \citenamefont
  {{Pupasov-Maksimov, A. M.}}(2014)}]{der14}%
  \BibitemOpen
  \bibfield  {author} {\bibinfo {author} {\bibnamefont {{Deriglazov, A. A.}}}\
  and\ \bibinfo {author} {\bibnamefont {{Pupasov-Maksimov, A. M.}}},\ }\href
  {\doibase 10.1140/epjc/s10052-014-3101-2} {\bibfield  {journal} {\bibinfo
  {journal} {Eur. Phys. J. C}\ }\textbf {\bibinfo {volume} {74}},\ \bibinfo
  {pages} {3101} (\bibinfo {year} {2014})}\BibitemShut {NoStop}%
\bibitem [{\citenamefont {Bures}(1969)}]{Bures1969}%
  \BibitemOpen
  \bibfield  {author} {\bibinfo {author} {\bibfnamefont {D.}~\bibnamefont
  {Bures}},\ }\href {\doibase 10.1090/S0002-9947-1969-0236719-2} {\bibfield
  {journal} {\bibinfo  {journal} {Trans. Am. Math. Soc.}\ }\textbf {\bibinfo
  {volume} {135}},\ \bibinfo {pages} {199} (\bibinfo {year}
  {1969})}\BibitemShut {NoStop}%
\bibitem [{\citenamefont {Uhlmann}(1976)}]{Uhlmann1976}%
  \BibitemOpen
  \bibfield  {author} {\bibinfo {author} {\bibfnamefont {A.}~\bibnamefont
  {Uhlmann}},\ }\href {\doibase 10.1016/0034-4877(76)90060-4} {\bibfield
  {journal} {\bibinfo  {journal} {Reports Math. Phys.}\ }\textbf {\bibinfo
  {volume} {9}},\ \bibinfo {pages} {273} (\bibinfo {year} {1976})}\BibitemShut
  {NoStop}%
\bibitem [{\citenamefont {Nielsen}\ and\ \citenamefont
  {Chuang}(2010)}]{Nielsen2010}%
  \BibitemOpen
  \bibfield  {author} {\bibinfo {author} {\bibfnamefont {M.~A.}\ \bibnamefont
  {Nielsen}}\ and\ \bibinfo {author} {\bibfnamefont {I.~L.}\ \bibnamefont
  {Chuang}},\ }\href@noop {} {\emph {\bibinfo {title} {{Quantum computation and
  quantum information}}}}\ (\bibinfo  {publisher} {Cambridge University
  Press},\ \bibinfo {year} {2010})\BibitemShut {NoStop}%
\bibitem [{\citenamefont {Sommers}\ and\ \citenamefont
  {Zyczkowski}(2003)}]{Sommers2003}%
  \BibitemOpen
  \bibfield  {author} {\bibinfo {author} {\bibfnamefont {H.-J.}\ \bibnamefont
  {Sommers}}\ and\ \bibinfo {author} {\bibfnamefont {K.}~\bibnamefont
  {Zyczkowski}},\ }\href {\doibase 10.1088/0305-4470/36/39/308} {\bibfield
  {journal} {\bibinfo  {journal} {J. Phys. A. Math. Gen.}\ }\textbf {\bibinfo
  {volume} {36}},\ \bibinfo {pages} {10083} (\bibinfo {year}
  {2003})}\BibitemShut {NoStop}%
\bibitem [{\citenamefont {Olver}\ \emph {et~al.}(2010)\citenamefont {Olver},
  \citenamefont {Lozier}, \citenamefont {Boisvert},\ and\ \citenamefont
  {Clark}}]{NistHB}%
  \BibitemOpen
  \bibinfo {editor} {\bibfnamefont {F.~W.~J.}\ \bibnamefont {Olver}}, \bibinfo
  {editor} {\bibfnamefont {D.~W.}\ \bibnamefont {Lozier}}, \bibinfo {editor}
  {\bibfnamefont {R.~F.}\ \bibnamefont {Boisvert}}, \ and\ \bibinfo {editor}
  {\bibfnamefont {C.~W.}\ \bibnamefont {Clark}},\ eds.,\ \href@noop {} {\emph
  {\bibinfo {title} {{NIST Handbook of Mathematical Functions}}}}\ (\bibinfo
  {publisher} {Cambridge University Press},\ \bibinfo {address} {New York,
  NY},\ \bibinfo {year} {2010})\BibitemShut {NoStop}%
\bibitem [{\citenamefont {Weisstein}(2015)}]{MathWorld}%
  \BibitemOpen
  \bibfield  {author} {\bibinfo {author} {\bibfnamefont {E.~W.}\ \bibnamefont
  {Weisstein}},\ }\href
  {http://mathworld.wolfram.com/GegenbauerPolynomial.html} {\enquote {\bibinfo
  {title} {{Gegenbauer Polynomial}},}\ } (\bibinfo {year} {2015})\BibitemShut
  {NoStop}%
\end{thebibliography}%
\end{document}